*Review*

# A Survey on High-Capacity OFDM-based Passive Optical Networks

**Mohammad Ghanbarisabagh** [1], **Gobi Vetharatnam** [2,*], **Elias Giacoumidis** [3], **Hossein Rouzegar** [4] and **Nasreddine Mallouki** [5]

[1] Department of Electrical Engineering, Faculty of Electrical Engineering and Computer Sciences, Islamic Azad University North Tehran Branch, Hakimieh, Tehran, Iran; m.ghanbarisabagh@iau-tnb.ac.ir, m.ghanbarisabagh@gmail.com

[2] Department of Electrical and Electronic Engineering, Lee Kong Chian Faculty of Engineering and Science, University Tunku Abdul Rahman, Malaysia; gobiv@utar.edu.my

[3] Dublin City University, School of Electronic Engineering, Radio and Optical Communications Lab, Glasnevin 9, Dublin, Ireland; elias.giacoumidis@dcu.ie

[4] Department of Electrical and Computer Engineering, Babol Noshirvani University of Technology, Babol, Iran; h.rouzegar@yahoo.com

[5] Sys'comLab, ENIT, BP.37 Le Belvédère 1002Tunis, Tunisia; mallouki_nasreddine@yahoo.fr

* Correspondence: m.ghanbarisabagh@iau-tnb.ac.ir, m.ghanbarisabagh@gmail.com

**Abstract:** The exponential growth of demand for high-speed internet and high bandwidth applications such as interactive entertainment in access networks mandates the requirement for higher optical signal speeds. On the other hand, since cost and energy efficiency should be concurrently preserved in access networks, passive optical networks (PONs) have emerged as a breakthrough solution. Currently, the 10 Gb/s (10GE-PON) standard which employs time division multiplexing is not sufficient to fulfill users' requirements, and therefore, more advanced and spectrally efficient modulation formats are required. A potential solution is the implementation of hybrid wavelength division multiplexing (WDM) with orthogonal frequency-division multiplexing (OFDM), in which a specific wavelength and electronic subcarriers are assigned to each optical network unit according to users' bandwidth requirements. This survey shows that future-proof PONs should be supported by an adaptively modulated WDM-OFDM architecture to maximize signal capacity and transmission-reach in PONs, while the employment of reflective semiconductor optical amplification and remodulation can potentially prevent the employment of an additional light source.

## 1. Introduction

Connecting the service provider central offices (COs) to residential and business subscribers is done by the access network, also known as the "first-mile network". Residential subscribers request high bandwidth first-mile access solutions with media-rich services. The leading broadband access solutions that have been extended recently include digital subscriber line technologies (DSL and ADSL) which ultimately enable 1.5 Mb/s and 128 Kb/s of downstream bandwidth and upstream bandwidth, respectively. Furthermore, due to signal distortion the distance between DSL subscribers and a CO must not exceed 18,000 ft. Despite some DSL modifications including very-high-bit-rate DSL (VDSL) which enables up to 50 Mb/s of downstream bandwidth, these technologies are limited to quite short distances. For instance, the highest distance of VDSL support area is limited to 1,500 ft. Compared to traditional copper-based networks; passive optical networks (PON) enable much higher bandwidth in the access network. PON is an optical fiber-based network which serves as a point-to-multipoint network architecture without using any active expensive component (e.g. optical amplifier). In PONs, via one or multiple 1: N optical splitters, an optical line terminal (OLT) at the CO can be connected to many optical network units (ONUs) at remote nodes (RN). The connection between OLT and ONU as a network is passive. PONs exploits one wavelength for each downstream and upstream direction (from CO to end user and vice versa) and via coarse wavelength division multiplexing (CWDM) multiply optical signals are aggregated on the same fiber [2]. Some of the main advantages of PON include the least utilization of active facilities, decreasing cabling infrastructure, cheap maintenance, capability of being connected to TV and being scalable enough [3]. Nowadays, there is an important requirement for increasing the signal capacity by expanding the number of PON users, increasing video content capacity etc. Time division multiplexing access (TDMA)-WDM-PON uses different wavelengths to send TDM frames to several users and suggests high ONUs capacity. However, it requires great synchronization because it suffers from low transmission speed. To use the same bandwidth with no interference, optical code division multiple access-WDM-PON (OCDMAWDM-PON) applies encoded data in smaller chips [1]. Despite its higher bandwidth compared to customary copper-based access networks, a PON further increases its bandwidth when using WDM technology, allowing multiple wavelengths to be transmitted in both upstream and downstream directions. WDM-PON architectures date back to the middle of 1990s [2] and although they support increased ONU capacity, high security and enhanced transmission rate, they need passive optical devices which makes it infeasible [1].

Alternatively, optical OFDM has been proposed as a breakthrough approach for high spectral efficient modulation. Among them, OFDM-PON is the most popular one where low-bandwidth optical sub-channels are considered. Combining OFDM technology with WDM thus generating a hybrid OFDM-WDM-PON system [3] could lead to enormous capacity without considerably adding complexity. As shown in Figure 1, modulated OFDM subscribers can be transmitted to various ONUs through several wavelengths (OFDM-WDM-PON strategy).

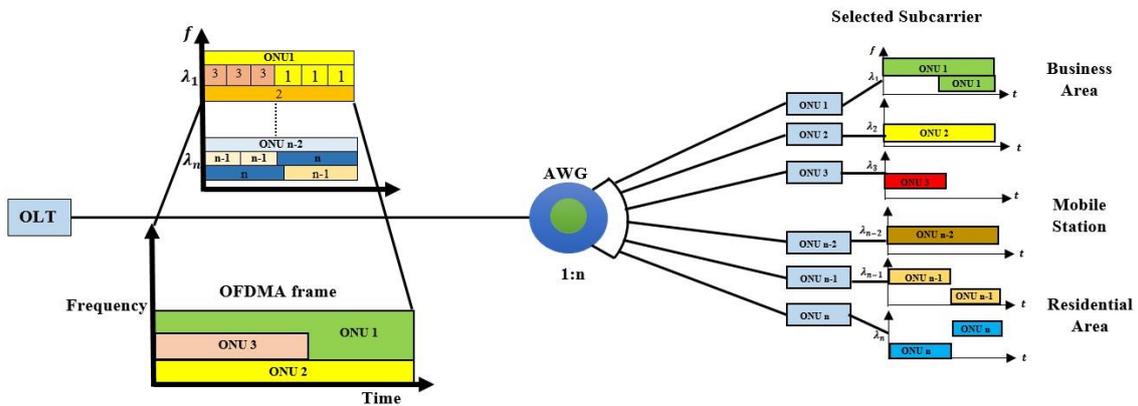

**Figure 1.** OFDMA-WDM-PON architecture for heterogeneous service delivery.

**2. Conventional PON architecture and terminology**

A typical PON architecture and terminology is depicted in Fig. 2. The main part of this architecture is the OLT, the optical network unit (ONU) and the optical distribution network (ODN). OLT serves as PON head-end which is generally placed in a CO. The ONU is generally located at the subscriber's premises, while the ODN includes optical power splitters and the transmission fiber and is located outside the plant (for example ducts and poles).

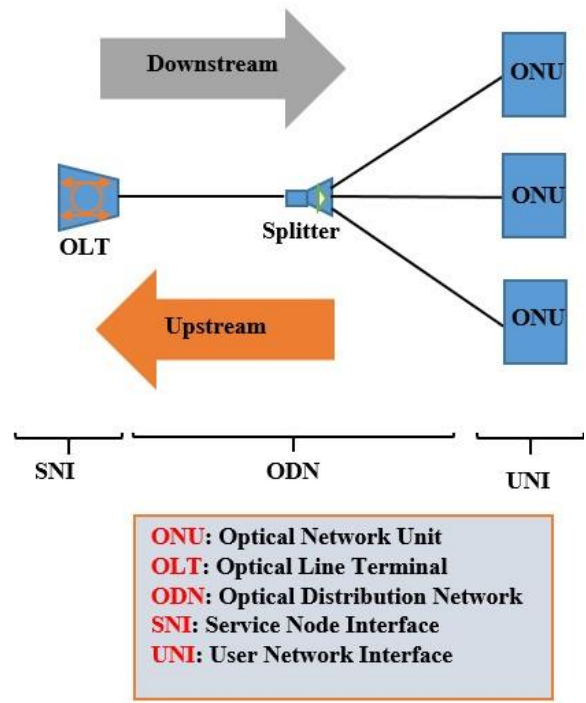

**Figure 2.** PON Architecture and Terminology.

The PON standards are contributed to the Full-Service Access Network (FSAN) and the International Telecommunication Union (ITU-T). According to FSAN/ITU-T, the next-generation (NG)-PONs are composed of two phases: NG-PON1 (mid-term promotions in PON networks) and NG-PON2 (long-term upgrades in PON evolution). NG-PON1 greatly requires to be integrated with the extended Gigabit PON (GPON) while reusing the outside plant. Some tests were performed in the recent Verizon field trials to satisfy these requirements. On the other hand, since ODNs usually occupy 70% of the whole investments, compatibility with deployed networks is essential for the NG-PON evolution. After such consideration, the only difficulty for migrating from GPON to NG-PON1 is the maturity of the "industrial chain". Although NG-PON1 has vivid purposes and vast progress, NG-PON2 has many candidate technologies to be considered. More specific, NG-PON2 technology should have good scalability, flexibility, reliability, higher bandwidth, and be more power and cost efficient than NG-PON1. It is finally anticipated that NG-PON2 would enable multi-service infrastructure thus allowing network operators to unify different service platforms to a single one, as depicted in Fig. 3.

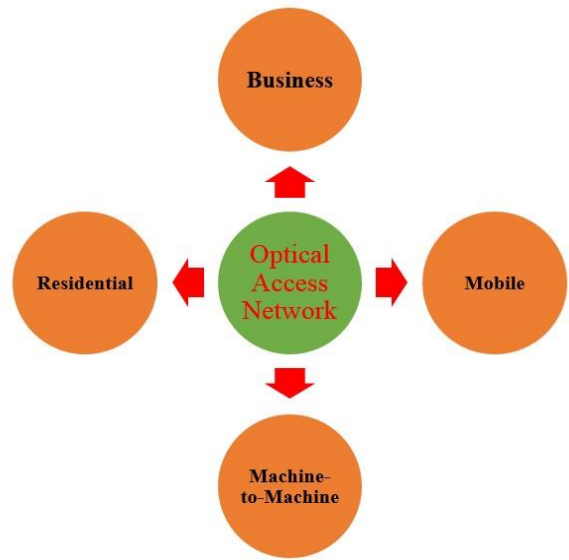

**Figure 3.** Future optical access networks are expected to be truly multi-service.

## 3. 3. WDM-PON

Since a WDM-PON supports different wavelengths over the same fiber infrastructure, being intrinsically clear to channel bit-rate and having no power-splitting losses, WDM-PON is scalable. Some other advantages of WDM-PONs are the high bandwidth with good data service, great splitting ratio, developed transmission access, collected backhaul for traffic, simple structure, improved end-user privacy and transparent protocol. Despite the aforementioned good features, WDM-PONs are not economically viable since they need an optical transmitter in the customer ONU to produce an optical carrier source which is accurately in alignment with a specially distributed WDM grid. To overcome such limitation, cheap light sources in the OLT are employed which are spectrally sliced. Therefore, the upstream data can be easily transmitted leading to colorless ONUs. Such light sources are typically composed of Fabry-Perot (FP) lasers or even light-emitting diodes (LEDs) [6]. However, the most conventional way to create a WDM-PON is by utilizing a distinct wavelength channel from the OLT to each ONU, for both downstream and upstream directions, as illustrated in Fig. 4.

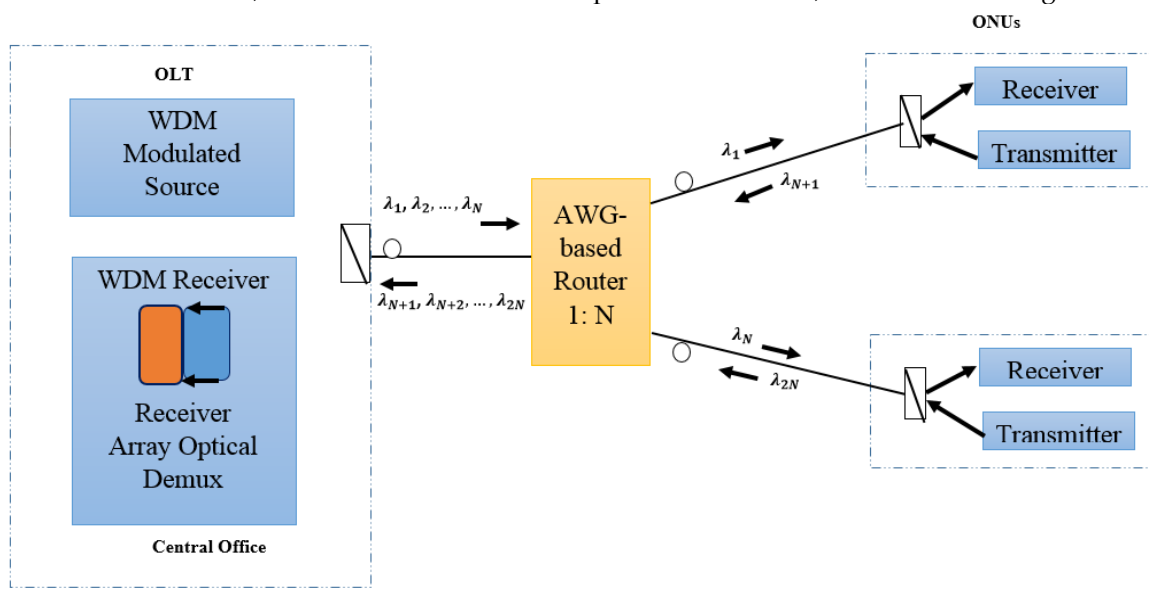

**Figure 4.** Conventional simplified WDM-PON architecture.

It should be noted that WDM-PONs typically requires adjustable bandwidth besides flexible number of users [2]. A WDM-PON has harnessed many devices such as array waveguide grating (AWG), WDM filters, reflective semiconductor optical amplifier (RSOA) and FP lasers. Such devices have led to many functional commercial products in business and wireless/wireline backhaul markets [7]. On Table 1, we show the transceiver and RN option for a WDM-PON.

**Table 1.** Transmitter, receiver and remote node (RN) options.

| | |
|---|---|
| **Transmitter option** | Wavelength-specified source: distributed feedback lasers (DFBs), vertical-cavity surface-emitting laser (VCSEL)-diodes, tunable lasers |
| | Multiple-wavelength source: mode-locked lasers, gain-coupled DFB LD array, chirped-pulse WDM |
| | Wavelength-selection-free source: spectrum-sliced source, injection-locked laser |
| | Shared source: SOA, external modulator |
| **Receiver options** | Photodiodes |
| | Recovery circuits |
| **Remote node options** | Power splitter, passive wavelength router |

Since each ONU is critical and expensive, many researchers have suggested removing optical sources from ONU. On the other hand, a slight deviation of one channel from the specified wavelength leads to the decay of both the channel and its neighbor ones. To solve this problem, all optical sources can be obtained from the OLT and the obtained unmodulated optical sources should be modulated by the ONU. Modulation of just a sectional temporary region for downstream and letting the remaining unmodulated region for upstream leads to a solution in which one wavelength channel is capable of being exploited in both directions. This option permits the employment of an external modulator and a SOA, i.e. the ONU splits the downstream signal and a part of that is sent to an external modulator. Modulation of this signal at high speed could make it suitable for upstream transmission. Another alternative is using an adjustable laser at the OLT, where for instance if the wavelength of the LD changes it can access each ONU. In this way, downstream data spends only half of the time while upstream spends the remaining half using an external modulator. Since such configuration may lead to a round-trip signal loss of the shared source and change of the external modulator output from the input signal polarization, a consideration of power margin and polarization is essential, i.e. the direction of the electric field that normally changes in a standard single-mode optical fiber (SSMF). It should be noted that a problem for the practical use of a modulator at each ONU is its cost. To compensate the round-trip signal loss, an RSOA is suggested to be applied as a shared source, as depicted in Figure 5. However, using a SOA the total cost of the PON system is still a problem which needs to be addressed before being considered as a potential commercial product [2].

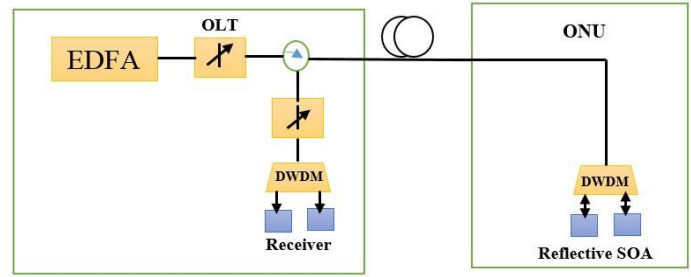

**Figure 5.** Shared source using a reflective SOA (RSOA).

At the receiver side, a photodetector (PD) and its participant electronics for signal recovery are typically employed. Positive-intrinsic-negative (PIN) and avalanche photodiode (APD) organize common PDs. Based on the essential sensitivity, PDs have different applications. Pre-amplifier, main amplifier, clock and data recovery circuits (CDRs) form the electronic parts which are dependent on the used protocol for each wavelength. Each WDM-PON receiver can also have a different configuration because of the different bandwidth requirements per wavelength. Utilization of wavelengths different from downstream in integral multiples of the free spectral range (FSR) of the AWG by upstream transmitters leads to the same AWG output port assigned for both upstream and downstream transmission, as depicted in Figure 6(a). This configuration employs a CWDM filter at the ONU to distinct the two signals. However, when the upstream shares the same wavelength with the downstream shared source, it is needed that the two different output ports be assigned to an ONU and a 2×N AWG should be utilized at the RN, as illustrated in Figure 6(b).

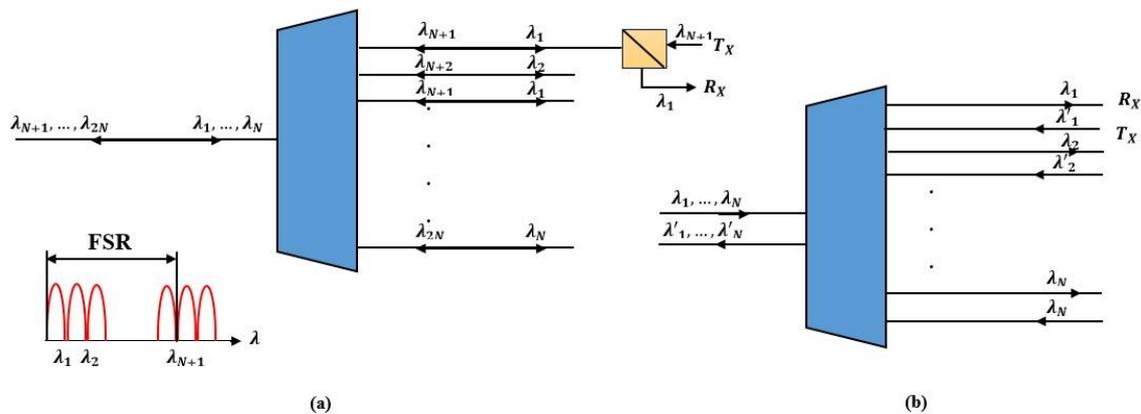

**Figure 6.** RN based on the cyclic wavelength property of the AWG: (a) bidirectional transceiver at the ONU, and (b) unidirectional transceiver at the ONU.

For the downstream the wavelengths are multiplexed to the ONUs while in the upstream the wavelengths are demultiplexed from the ONU by the first periodic AWG located at the CO. A numerous free spectral range (FSR) of the AWG separates the downstream and upstream wavelengths which are allocated to each ONU. This causes both wavelengths to be guided in and out of the same AWG port which is connected to the targeted ONU. As shown in Figure 7, $\lambda_1, \lambda_2 \ldots \lambda_N$ are downstream wavelengths assigned for ONU1, ONU2, and ONUN, respectively. In a similar way, $\lambda_1', \lambda_2' \ldots \lambda_N'$ are the upstream wavelengths from ONU1, ONU2 and ONUN, respectively which are headed to the CO. The wavelength channels in a WDM PON usually cover 100 GHz (0.8 nm) of frequency space. Dense WDM-PON (DWDM) systems on the other hand utilize channel spacing of 50 GHz or even less. Despite the physical point-to-multipoint (P2MP) topology of a WDM

PON, the CO and each ONU are connected by logical point-to-point (P2P) connections. At ONU N, downstream signals are received on $??_N$ and upstream signals are transmitted on $??_N$. Such ONU assigns specific capacity to these wavelengths.

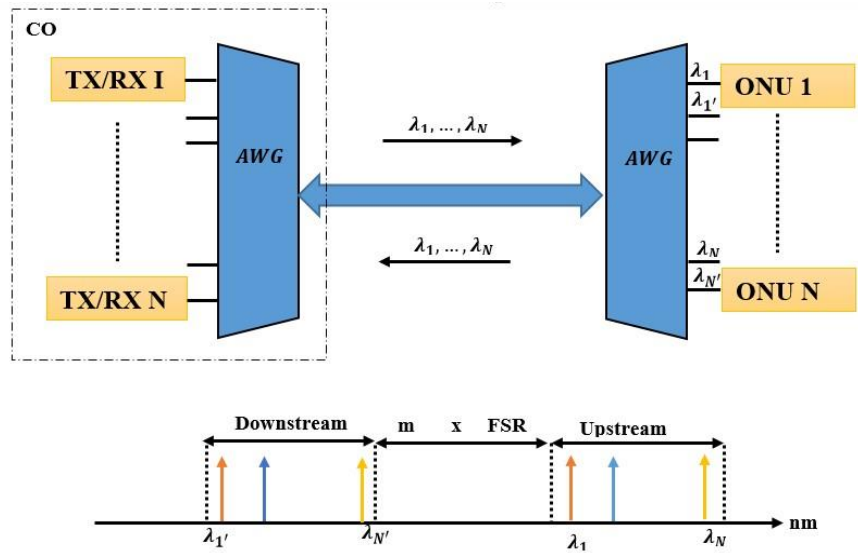

**Figure 7.** WDM-PON architecture. Inset: Allocation of upstream and downstream wavelength channels into two separate wavebands.

Another modern WDM-PON configuration includes the utilization of a fixed-wavelength laser array or a multi-frequency laser (MFL). Fig. 8(a) shows the 'broadcast-and-select' architecture in which all wavelengths in the downstream directions are transmitted via a passive 1:N splitter. Each ONU selects one of the downstream wavelengths by employing an individual filter, while another individual wavelength is utilized for the upstream direction. The 1:N coupler incorporates the upstream wavelengths passively, however, this structure leads to some losses from the passive splitter/combiner and the TDMA. Both the transmitter and receiver filters should be tunable to allow the usage of identical ONUs. Fig. 8 illustrates an AWG-based (arrayed waveguide grating) wavelength-routing PON. Here an AWG wavelength router is utilized instead of the passive splitter/combiner causing lower insertion loss (i.e. with no dependency to the number of wavelengths; AWGs have a typical insertion loss of 5 dB). Furthermore, no wavelength selective (individual) receivers (denoted as 'Rx') are needed so it leads to simplification of the ONUs.

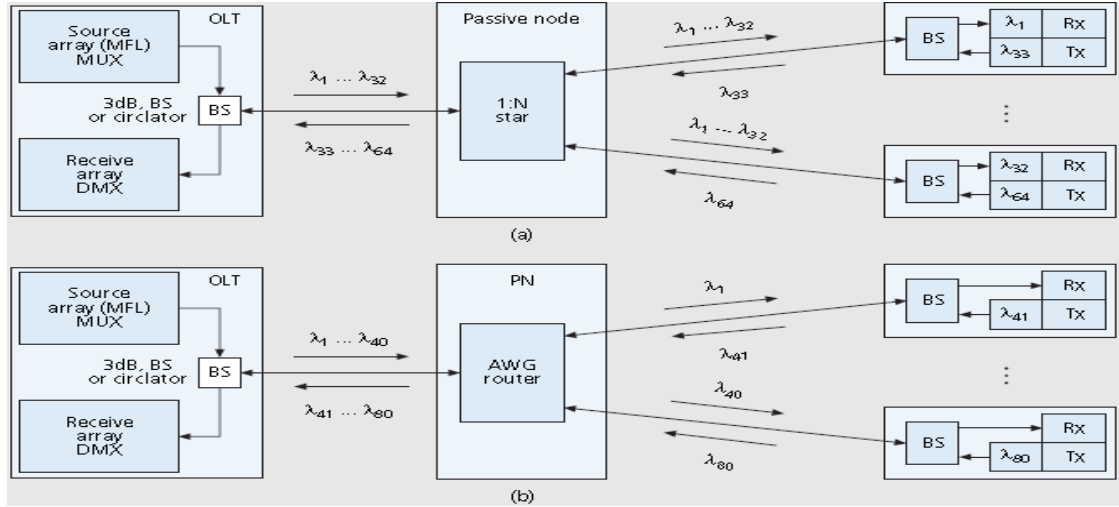

**Figure 8.** Alternative WDM-PON architectures: a) 'broadcast-and-select' WDM-PON with splitter/combiner in passive node; b) AWG-based wavelength routing PON (MFL: multi-frequency laser; BS: band splitter) [8].

Finally, the extended-reach dense WDM-PONs (ER-DWDM-PONs) keep all the main properties related to conventional WDM-PONs while converting optical access networks and metropolitan area networks. This consequently leads to a decrease in the number of equipment interfaces and network elements. ER-DWDM-PONs is considered as an ambitious tactic for high-capacity next-generation PONs (NG-PONs). The most important challenges for practical development of ER-DWDM-PONs are cost-efficiency and flexibility. The key solution to provide cost-effectiveness in ER-DWDM-PONs is intensity modulation with direct detection (IMDD) because of its simplicity such as using one PD and directly modulated DFB lasers (DMLs). DMLs are in particularly preferred to other intensity modulators since they provide cost and power consumption reduction, compactness, high output power and relatively low driving voltage [9, 10].

**4. OFDM-based PON**

OFDM appertains to a larger class of multicarrier modulation (MCM), where many lower rate subscribers carry the data information as shown in Fig. 9. More specific, in OFDM the signal is split into several narrowband channels (also known as subcarriers) at different frequencies carrying low bandwidth signal capacity. Being robust against channel dispersion and simple phase and channel estimation in an environment varying with time are two of the main benefits of OFDM [11]. Some of its fundamental disadvantages include the high peak-to-average power ratio (PAPR) and the sensitivity to frequency offset and phase noise. While not commercialized yet in optical domain, OFDM has been standardized in wireless communications by Wi-Fi, GSM, WiMAX and LTE operating at microwave frequency band (2-4 GHz). A transmitted signal of MCM/OFDM namely s(t) is obtained by:

$$s(t) = \sum_{i=-\infty}^{+\infty} \sum_{k=l}^{N_{sc}} c_{ki} s_k(t - iT_s), \tag{1}$$

$$s_k(t) = \prod(t) e^{j2\pi f_k t} \tag{2}$$

$$\prod(t) = \begin{Bmatrix} 1, & (0 < t < T_s) \\ 0, & t \leq 0, t \geq T_s \end{Bmatrix} \tag{3}$$

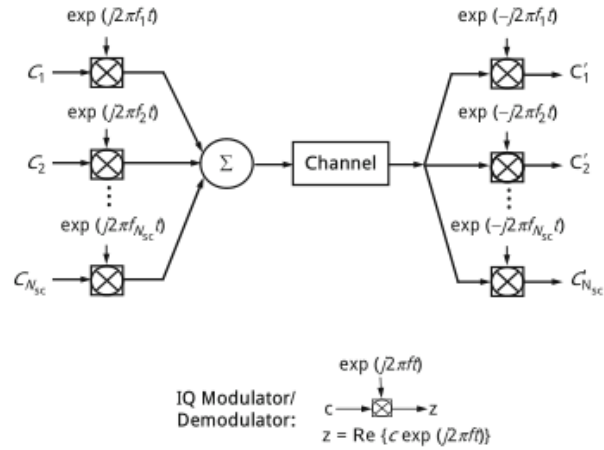

**Figure 9.** Conceptual diagram for a generic multicarrier modulation system.

A frequency-selective channel can be transformed to a parallel collection of frequency flat sub channels by OFDM.

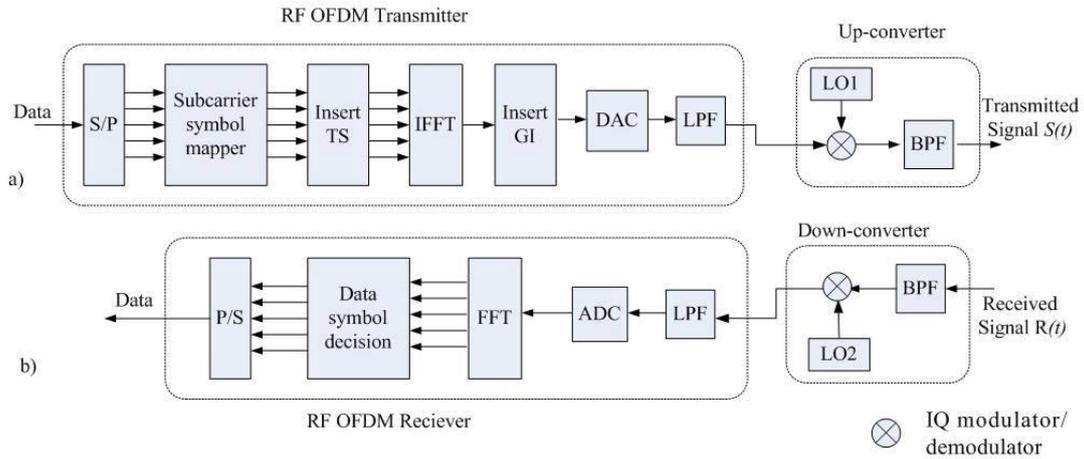

**Figure 10.** Building blocks of OFDM system for (a) transmitter and (b) receiver side.

The low-pass equivalent OFDM signal is expressed as:

$$X(t) = \sum_{k=0}^{N-1} X_k e^{j2\pi kt/T}, \qquad 0 \leq t < T \qquad (4)$$

This is equivalent to the discrete Fourier transform (DFT). In Eq. (4) $X_k$ corresponds to data symbols which are a sequence of complex numbers representing different signal modulation formats such as binary phase-shift keying (BPSK), quaternary PSK (QPSK) or quadrature amplitude modulation (QAM) baseband. Where N is the number of OFDM subcarriers and T is the symbol rate. 1/T is the subcarrier spacing which makes the subcarriers orthogonal per symbol period. The total sequence of the OFDM symbols is given by [12]:

$$S(t) = \sum_{k=-\infty}^{+\infty} X(t - kT) \qquad (5)$$

Since different frequencies are used for parallel data transmission in OFDM, this leads to a much longer symbol period compared to single-carrier modulation. In OFDM the residual ISI is removed by utilizing a cyclic prefix (CP) in time domain which is a type of guard interval (GI) and contains no useful information. [13]. It was suggested that CP can compensate both ISI and inter-carrier interference (ICI) induced by the channel dispersion [11].

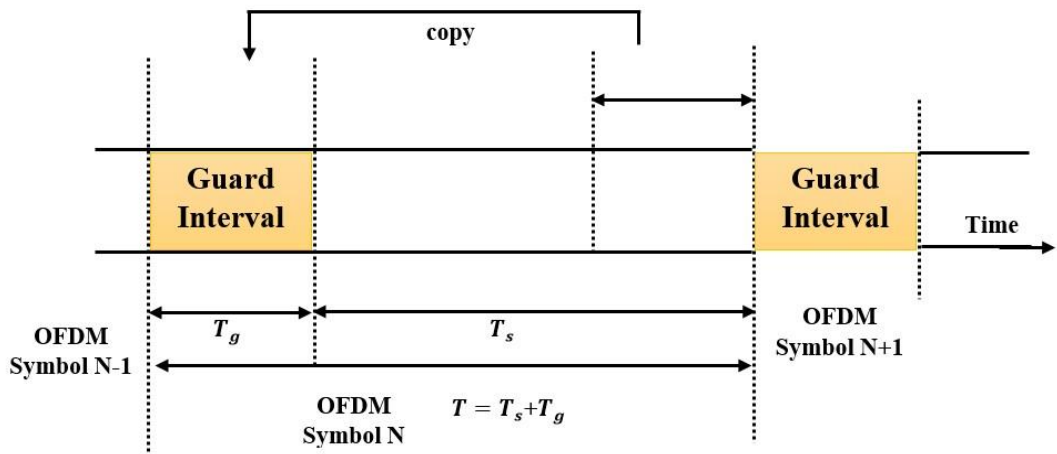

**Figure 11.** Insertion of cyclic prefix in OFDM symbols.

The OFDM signal under the presence of a CP length (⊚G) is expressed as,

$$X(t) = \sum_{k=0}^{N-1} X_k e^{j2\pi kt/T}, \qquad -\Delta G \leq t < T \qquad (6)$$

The waveform in the "CP area" is exactly an identical copy of that in the DFT window, being time-shifted forward by $t_s$, as depicted in Fig. 11. The OFDM signal including the CP upon reception is presented in Fig. 12 [11], in which the condition for ISI-free OFDM transmission is given by $t_d < a??G$.

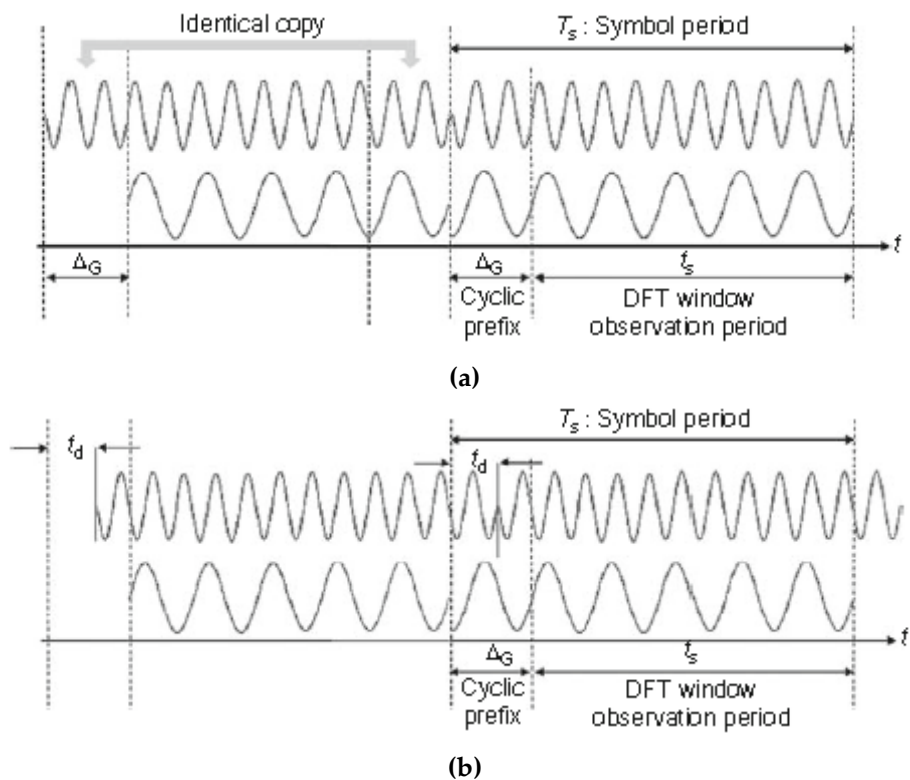

**Figure 12.** Time domain OFDM waveforms with a cyclic prefix at (a) transmitter and (b) receiver side.

For the proper recovery of the OFDM information, two main methods have been suggested: (1) DFT window synchronization, which is done by selecting an appropriate DFT window, and (2) channel estimation or subcarrier recovery which is the estimation of the phase shift for each subcarrier [11]. OFDM modulation and demodulation is traditionally done using the fast Fourier Transform (FFT) and its inverse form, which is an efficient implementation of DFT from a computational point of view [12].

Analytically, OFDM is expressed by utilizing overlapped yet orthogonal signal sets. The base of this orthogonality is a direct correlation between any two subcarriers as shown below:

$$\delta_{kl} = \frac{1}{T_s}\int_0^{T_s} s_k s_l^* dt = \frac{1}{T_s}\int_0^{T_s} e^{j2\pi(f_k-f_t)t} dt = e^{j\pi(f_k-f_t)T_s}\frac{sin(\pi(f_k-f_t)T_s)}{\pi(f_k-f_t)T_s} \quad (7)$$

By satisfying the condition $f_k - f_l = m\frac{1}{T_s}$, the two subcarriers will be orthogonal to each other [11]. One major drawback of OFDM as stated before is its high sensitivity to frequency offset and phase noise [13] partly due to its spectrum where each OFDM subcarrier has prominent frequency side lobes as depicted in Figure 13.

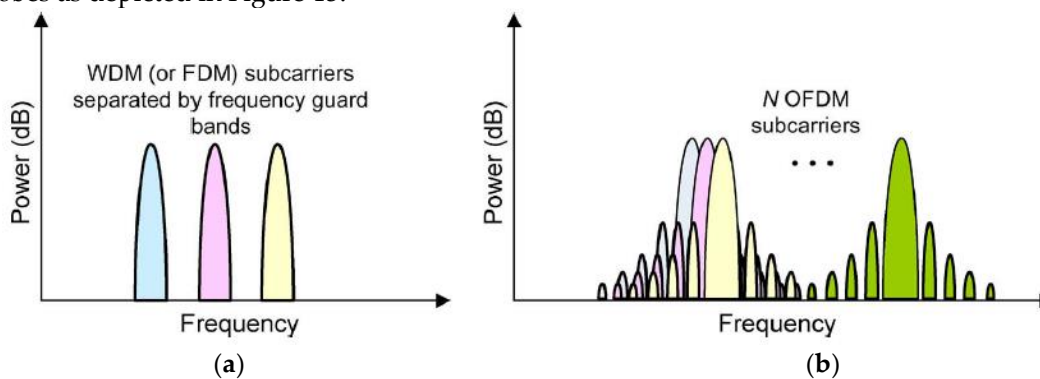

**Figure 13.** Spectrum of (a) WDM or FDM signals and (b) subcarrier-based OFDM [13].

The second drawback of OFDM, as also mentioned previously, is its high PAPR which causes inter-subcarrier intermixing distortion especially when a high number of subcarriers are employed. This can be explained as follows: the summation of N sinusoids in OFDM lead to superposition due to the inverse FTT (IFFT) process at the transmitter-end, and hence large power peaks are generated by some combinations of these sinusoids. These peaks cause difficulties at different stages of an OFDM system, including the word length of IFFT/FFT, the digital-to-analogue/analogue-to-digital converter (DAC/ADC), and most importantly the high-power amplifier (HPA) which is designed to control irregular coincidences of large peaks. For the latter case, HPA may experience both out-of-band (OOB) and in-band (IB) distortions, thus forbidding it to operate in linear region (saturation region operation). The PAPR of an OFDM signal, x, can be expressed as follows:

$$PAPR = \frac{max|x(t)|^2}{E[|x(t)|^2]} \quad (8)$$

Several PAPR reduction techniques have been proposed which are categorized as follows: clipping, scrambling, and adaptive pre-distortion, convex optimization, coding, and pre-coding based techniques [15]. Because of its very-high spectral efficiency and simplicity compared to single-carrier modulation, OFDM has attracted a lot of attention in optical domain for both access and core optical networks [18]. In optical field, the first optical OFDM report was reported back in 1996 [11]. OFDM for optical communications is categorized into two fundamental divisions: coherent detection (CO-OFDM) and direct detection (DD-OOFDM), where (optional) data could be also be transmitted in two polarizations of the optical fiber to double the transmission capacity [19, 20]. In Fig. 14, a generic CO-OFDM block-diagram is illustrated in which five main functional blocks are involved: (1) radio frequency (RF) OFDM transmitter, (2) RF-to-optical (RTO) up-converter, (3) optical channel, (4) optical-to-RF (OTR) down-converter, and (5) RF OFDM receiver.

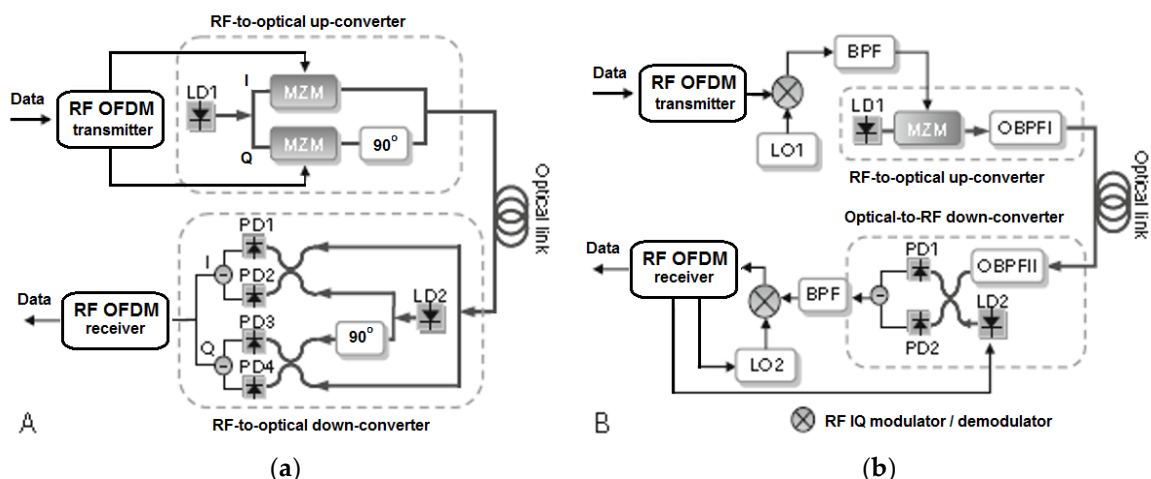

**Figure 14.** A CO-OFDM system in (a) direct up-/down-conversion architecture and (b) intermediate frequency architecture.

While previously we discussed the general aspects of OFDM technology and its potential to be brought into optical domain, now we shall deliberate the OFDM implementation scenarios for PONs. Generally speaking, more challenging than the downstream direction is the upstream multipoint-to-point architecture of PONs. The upstream direction is even more challenging for OFDM-access (OFDMA)-PONs, in which orthogonality is hard to be maintained among subcarriers for various reasons so that data can be efficiently recovered. One of the main reasons for OFDM signal degradation in PON upstream direction is the optical beat interference (OBI): the origin of OBI is that multiple optical sources in the upstream direction are presented in a PON system. Yet, concurrent transmission of two or more ONUs/lasers on the same channel with a small frequency separation (typically of a few GHz) leads to mixing of the optical field during the photo-detection process. More specific, beat signals are generated in the photocurrent (cross-mixing terms at the difference frequencies related to each pair of optical fields) by the square law nature at the OLT. These interference terms can overlap an active subcarrier channel, causing increase of additional noise which is introduced as OBI [16]. Giacoumidis et al [16] proposed a cost-effective IMDD-based multi-channel OFDM-PON system at a signal bit-rate up to ~20 Gb/s per channel where a combination of adaptive loading algorithms (e.g. bit-and-power loading), clipping and thermal detuning (TD) can effectively tackle the PAPR and OBI obstacles. In more detail, it was revealed that a very efficient PAPR reduction technique is the bit-loading algorithm, in which different modulation format levels (e.g. 16-64 QAM) are adjusted according to individual subcarrier signal-to-noise ratio (SNR) and total signal bit-error-rate (BER). This occurs because bit-loading reduces the probability of independently modulated OFDM subcarriers to be added up coherently by the IFFT [17]. On the other hand, TD along with bit-loading is considered a very effective and relatively simple approach to suppress the OBI effect.

While some of advantages of OFDM over single-carrier modulation include higher spectral efficiency, simple chromatic dispersion compensation and channel estimation/equalization, commercialization has not been considered yet since more substantial benefits are required. Adaptive OFDM has been proposed to exploit OFDM to its full potential to considerably outperform single-carrier modulation in terms of signal capacity, network flexibility, and performance robustness. Based on the characteristics of a given transmission link, in this technique a high (low) signal modulation format (e.g. 16-64 QAM) is exploited on a subcarrier that has a high (low) SNR [10]. In a similar manner the power levels can also be adjusted according to individual subcarrier SNR. The first implementation is called adaptive modulation or bit-loading (BL), while the latter is referred to power-loading (PL). There is also an advanced option to combine both cases thus developing the bit-and-power loading (BPL) algorithm [10]. It should be noted that, if any subcarrier experiences a very low SNR, it should be discarded to prevent significant errors [10]. Adaptive OFDM typically significantly outperforms conventional OFDM utilizing a similar signal modulation format across all subcarriers [10], which is highly beneficially for ER-DWDM-OFDM-PONs. A typical transceiver block diagram of IMDD wavelength-division multiplexed (WDM) OFDM-based PON system using adaptive OFDM is shown in Fig. 15:

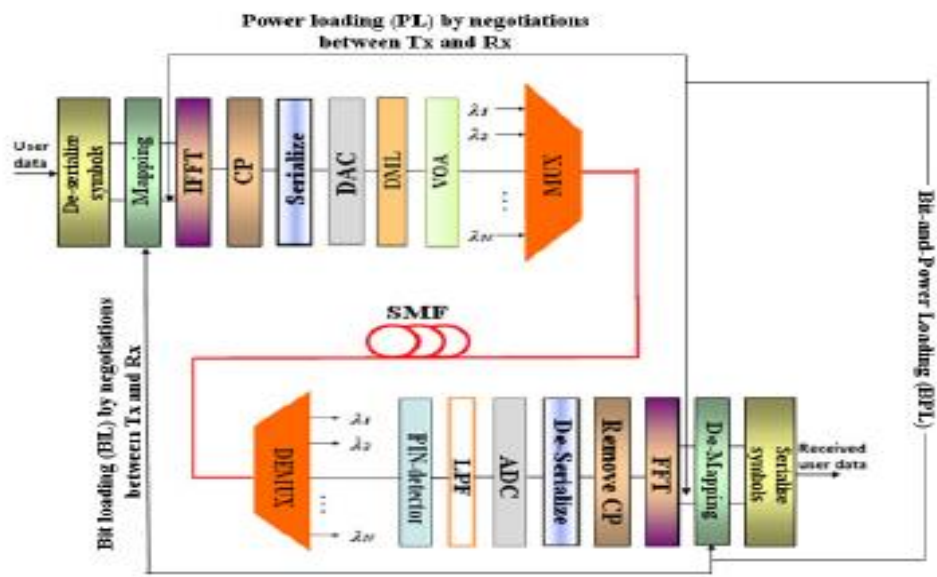

**Figure 15.** A DML-based IMDD wavelength-division multiplexed (WDM) OFDM PON system using bit-loading (BL), power-loading (PL) and bit-and-power loading (BPL) [21].

As seen from Fig. 15, in-line optical amplification and chromatic dispersion compensation is not necessary compared to state-of-the-art optical systems, thus showing adaptive OFDM's great potential for implementation in cost-efficient high-capacity PONs. In more detail, in Fig. 15 an electrical OFDM signal with a positive sign incorporated with a suitable DC bias current is utilized to directly drive a DML, from which an optical OFDM signal is generated at a determined wavelength. After the DML, the optical signal power at the required level is adjusted by a variable optical attenuator (VOA). These procedures are repeated by utilizing different incoming randomized data sequences to produce WDM channels with various wavelengths spaced at the desired frequency interval. A multiplexer (MUX) sums all the WDM signals, and afterwards the incorporated WDM OFDM signals are transmitted over a standard single-mode fiber (SSMF). At the receiver-side a de-multiplexer (DEMUX) separates the received WDM signals with a spectral bandwidth of half of the channel spacing. A square-law photo-detector detects each separated WDM channel and the inverse procedure is utilized to process the down-converted electrical signal. Finally, the data are recovered for each WDM channel and the BER is calculated following typical digital signal processing (DSP) de-modulation similarly to Ref. [10]. In contrast to conventional optical OFDM-PON systems, the aforementioned system includes additivity of OFDM subcarriers by negotiations between transmitter and receiver until the total system BER reaches the targeted values. More specific for each algorithm we consider:

- BL: In BL the signal modulation format is adjusted according to the system frequency response. More specific, a high (low) signal modulation format is applied on a subcarrier with a high (low) SNR [21]. The signal line rate of the wavelength channel is determined by the level of signal modulation format on individual subcarriers within an OFDM symbol for a particular wavelength channel. The rate of signal line for each WDM channel is calculated by:

$$R_j = \frac{r_{sj} \sum_{K=1}^{N_{sj}-1} n_{kj}}{2N_{sj}(1 + C_{pj})} \tag{9}$$

Here $j$ is the WDM channel index. $N_{sj} - 1$ is the total number of data-earing subcarriers in the positive frequency bins, $n_{kj}$ is the total number of binary bits which is carried by the k-th subcarrier within one symbol period, $r_{sj}$ is the sampling rate of an analog-to-digital converter (ADC)/digital-to-analog converter (DAC) employed in the j-th WDM channel. There is a deal between the transmitter and the receiver in the initial stage when establishing a transmission link which increases the signal capacity of the j-th WDM channel because of assigning the highest possible signal-modulation format on each subcarrier. It can be implemented by detecting the total BER of the j-th WDM channel, i.e., $BER_{Tj}$, and its commensurate subcarrier BER, i.e., $BER_{kj}$. These parameters are obtained as follows:

$$BER_{Tj} = \frac{\sum_{K=1}^{N_{sj}-1} En_{kj}}{\sum_{K=1}^{N_{sj}-1} Bit_{kj}} \quad (10)$$

$$BER_{kj} = \frac{En_{kj}}{Bit_{kj}} \quad (11)$$

Where the $En_{kj}$ is the whole number of errors that are detected in the entire data sequence adopted in the j-th WDM channel, and $Bit_{kj}$ is the whole number of transmitted binary bits of the data sequence adopted in the j-th WDM channel. $En_{kj}$ and $Bit_{kj}$ belong to the k-th subcarrier. It is notable that the signal line rate which can be calculated by Eq. (9) is viable just when the corresponding targeted $BER_{Tj} = 10^{-3}$ is satisfied.

- PL: All OFDM subcarriers take a maximum possible level of signal modulation format according to a total channel $BER_T$ of $\leq 10^{-3}$. As a practical example based on [9], a combination of advanced functionalities was considered for the experimental demonstration of the fastest ever 11.25 Gb/s real-time FPGA-based OOFDM transceivers which uses 64-QAM encoding/decoding with PL on each individual subcarrier. The implemented transceivers were entirely made from off-the-shelf electrical and optical components. Using a practical DML-based IMDD real-time end-to-end 64-QAM-optical OFDM system, a high electrical spectral efficiency of 5.625 bit/s/Hz was achieved over 25 km of standard and MetroCor SMFs with power penalties of 0.3 dB and −0.2 dB at a BER of $\leq 10^{-3}$ without in-line optical amplification and chromatic dispersion compensation.

- BPL: A graphical description of the BPL implementation procedure is shown below in Figure 16.

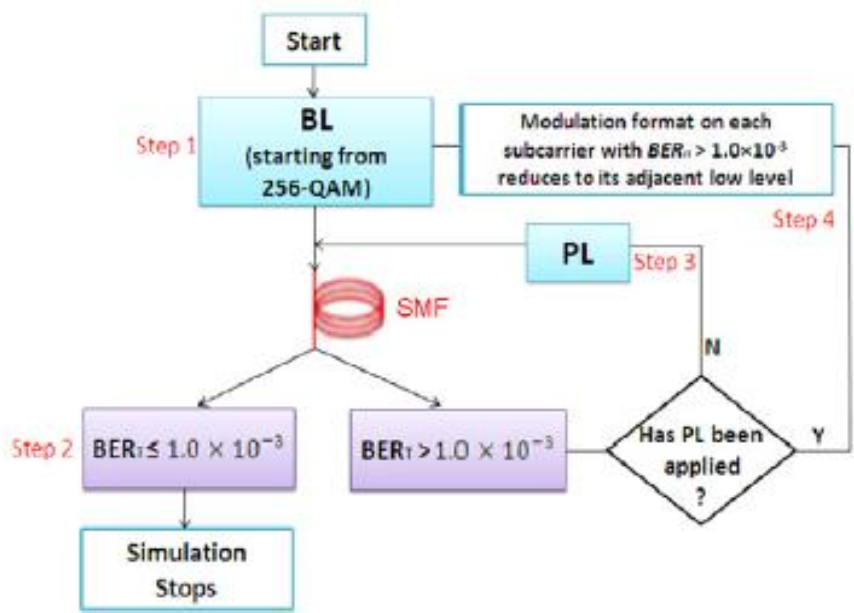

**Figure 16.** The BPL implementation procedure [21].

The generic bit/power loading optimization problem is described as follows:

$$max(R) = max\left(\sum_{k=1}^{N_s} R_k\right) \quad (12)$$

with the constraint of:

$$\sum_{k=1}^{N_s} P_i = P_o \tag{13}$$

where R is the composite data-rate, $R_k$ is the individual subcarrier data-rate, $P_O$ is the composite total power and $P_i$ is the individual subcarrier power. According to Eqs. (12) and (13), adjusting the order of modulation and the power of each subcarrier is possible. It should be noted that the optical channel is relatively "gentler" than wireless systems in which the subcarrier SNR varies by tens of decibels due to multipath fading. Hence, BPL algorithm is adopted differently in an optical OFDM system as it typically takes less time to converge (the negotiating-time between transmitter and receiver to reach the required BER).

As an example, we summarize the results of BL-optical OFDM for PONs as analyzed by [10]: The effects of cross-phase modulation (XPM) and four-wave mixing (FWM) are examined using 3 cases as shown below which is the transmission performance central WDM channel in its worst case:

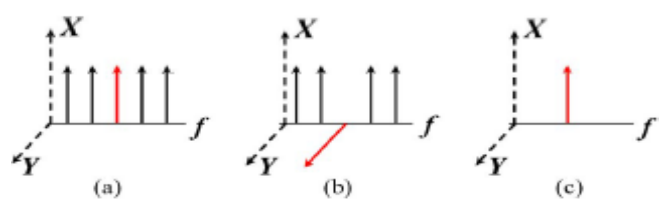

**Figure 17.** Polarization states of five WDM channels at the input facet of a WDM transmission system. (a) All x-polarized WDM channels, (b) 4 x-polarized and 1 y-polarized WDM channels, and (c) a single x-polarized channel.

According to Figure 17, Case (a) includes five x-polarized WDM channels. In this case, the XPM and FWM effects impact the central channel. For case (b), the y-polarization is assigned to the central channel, but the other four WDM channels keep being at x-polarization. This signal polarization arrangement has weaker impact of XPM on middle channel than in Case (a) by 1/3. Eventually, Case (c) consists of only one single x-polarized channel with no XPM and FWM effects. In this case, the nonlinearity penalty will come from self-phase modulation (SPM) only. In Fig. 18, results are shown [10] for Cases (b) and (c): in the WDM nonlinearity-limited performance region, BL-optical OFDM (also referred to as AMOOFDM from adaptively modulated optical OFDM) increases the signal bit-rate by a factor of > 1.3 while improving the optimum optical launch power to almost 5 dB, compared with conventional optical OFDM [10]. It also shows that XPM/FWM plays an important role on system transmission performance.

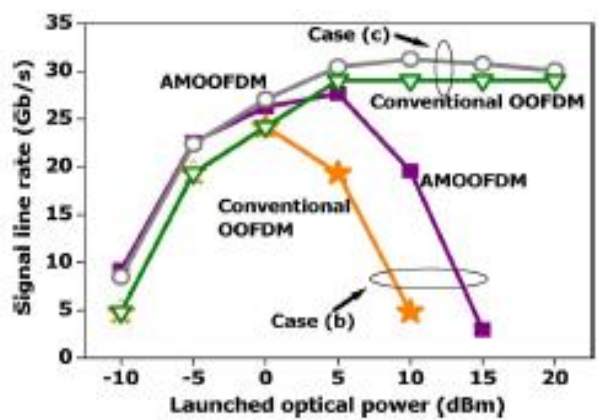

**Figure 18.** Transmission capacity versus optical launch power for IMDD AMOOFDM (bit-loaded optical OFDM) and conventional optical OFDM after 40 km of transmission [10].

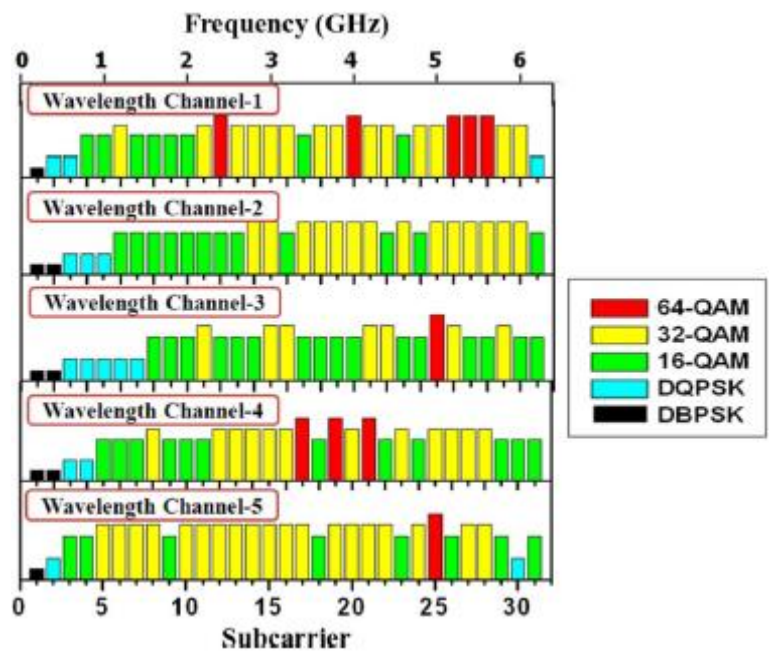

**Figure 19.** Distribution of signal modulation format through all the subcarriers of five WDM channels with the same polarization states. Cross-channel complementary modulation format mapping is exhibited [10].

In more detail, in Fig. 19 we show the adopted modulation format levels (DBPSK up to 64-QAM) on different subcarriers (which are 31 [10]) for all five wavelength channels using the bit-loading algorithm According to Fig. 19, wavelength channel 3 owns the lowest average signal modulation format level and the smallest transmission capacity, while higher average signal modulation format levels and larger transmission capacities belong to wavelength channel 1 and wavelength channel 5. This is a direct result from XPM and FWM among various WDM channels, in which the central channel experiences the strongest crosstalk effect. This effect decreases by the increase of frequency difference from the central channel. Moreover, the triangle-shaped signal spectral distortion across the entire WDM window leads to cross-channel complementary modulation format mapping, as depicted in Figure 18. This means that, for wavelength of channels 1 and 2, high-frequency subcarriers take relatively high signal modulation formats, while wavelength channel 4 and wavelength channel 5, high-frequency subcarriers take relatively low-signal-modulation formats. This cross-channel complementary nature only occurs when applying AMOOFDM. Essentially, AMOOFDM significantly reduces XPM and FWM, thus increasing the total signal capacity compared to conventional optical OFDM. Finally, it should be noted that subcarrier × subcarrier intermixing effect causes adaptation of low-modulation formats on the low-frequency subcarriers, while the effects of the DML-induced frequency chirp and system frequency response roll-off are the reason for low-modulation formats on the high-frequency subcarriers.

Optical OFDM has been proposed as a "future-proof" solution for next-generation PONs to increase the signal capacity in both downstream (DS) and upstream (US) directions. More specifically, it is expected optical OFDM to overcome current PON standards (e.g. G-PON, XG-PON of ≤10 Gb/s) reaching up to >40 Gb/s per user. An example of a proposed reference network architecture (RNA) using OFDM technology as reported from the EU FP7 project ACCORDANCE [22, 23] is shown below in Fig. 20, targeting the upcoming fiber-to-the-home (FTTH) service.

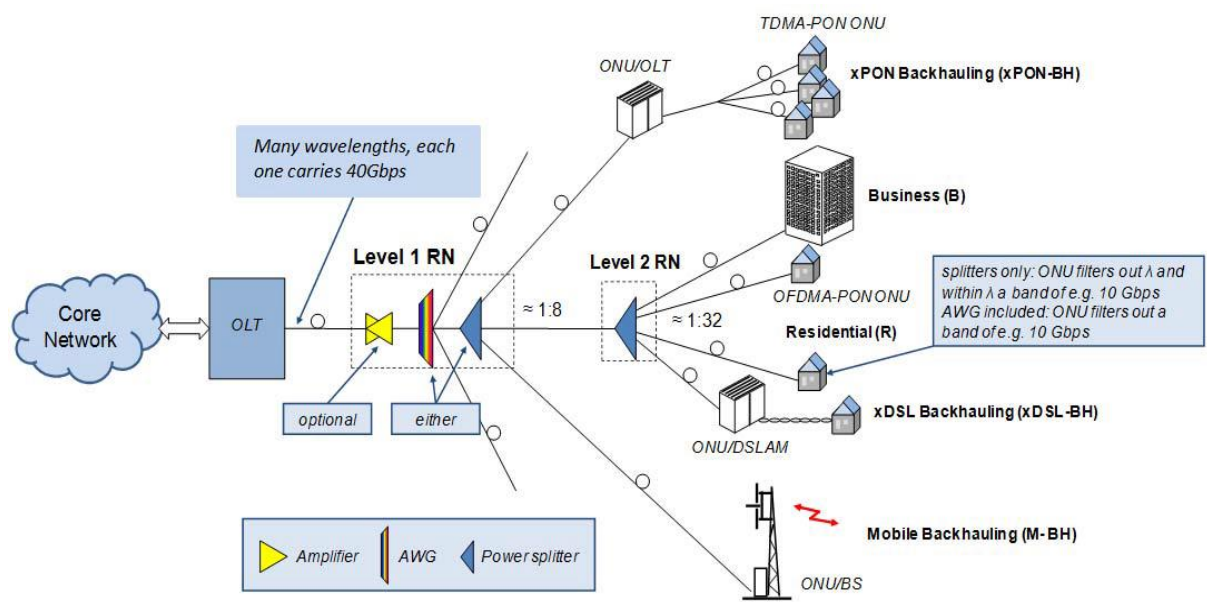

**Figure 20.** The ACCORDANCE reference network architecture (RNA) [22, 23].

Among different technologies suggested for OFDMA-PON, there are some solutions that can be considered as the best ones including: 1) intensity-modulation and direct-detection (IM/DD). There are two types of laser in IM/DD: a DML like a distributed feedback (DFB) laser or an external modulation laser (EML) like a DFB with a Mach-Zehnder modulator (MZM). This kind of modulation is referred to as intensity-modulation (IM). Exploiting a direct-detection (DD) in the receiver, the optical signal is detected by PIN photodiode. 2) RF-up conversion with direct-detection (RF/DD) and 3) RF-up conversion with coherent detection (RF/CO): In these two techniques, the complex OFDM baseband signal is up-converted to an intermediate frequency in the electrical domain by an electrical (RF) IQ modulator. A pair of RF mixers and local oscillators (LOs) with a 90-degree shift creates this IQ modulator. After the electrical IQ modulator, the MZM up-converts (UP) the signal to the optical domain. For detection, a simple photodiode and a coherent receiver with a pump laser and four photodiodes are utilized for DD and CO cases, respectively. Finally, all-optical solutions can also be considered such as: 4) optical IQ modulation with CO detection (IQ/CO); 5) orthogonal-band multiplexing (OBM) and 6) all-optical OFDM (All-O) [23]. A qualitative comparison of the aforementioned solutions is shown below in Fig. 21.

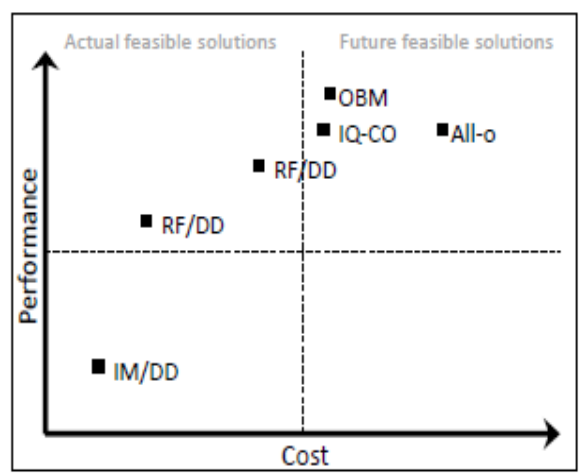

**Figure 21.** The ACCORDANCE Reference Network Architecture (RNA) [23].

In Fig. 22, the dominant optical OFDM transceiver solutions are shown: (1) intensity-modulation (IM) with (5) direct-detection (DD) or using a (6) coherent (CO) receiver; (2) optical frequency modulation (oFM) with either (3) RF-up transmitter (4) or optical IQ modulation in combination with either (5) or (6).

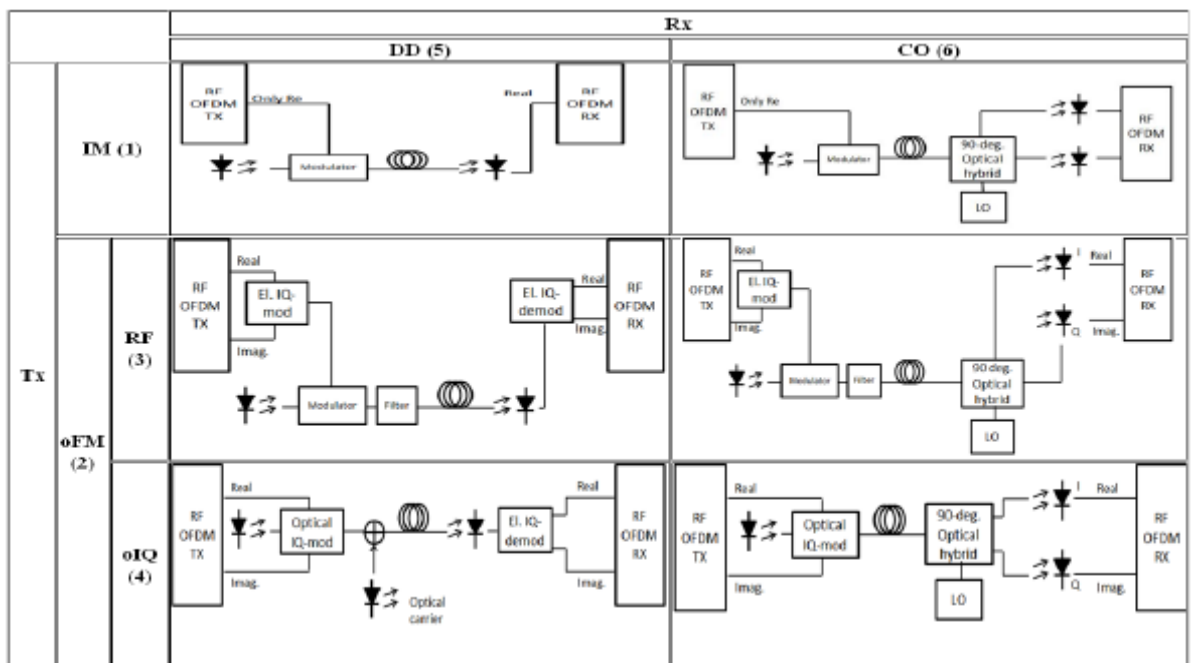

**Figure 22.** Optical OFDMA technology solutions block diagrams [22].

**5. Wavelength reused WDM-PONs using RSOAs**

Wavelength reused WDM-PONs DS wavelength reuse [6, 24] has been of an enormous research interest. In its structure, there is no need for any extra light source in the ONU but it requires the DS optical signal. For upstream data re-modulation in the ONU in wavelength reused WDM-PONs, an RSOA intensity modulator receives some part of downstream optical signal. The cost-effectiveness and wavelength control functionalities of WDM-PONs can be enhanced by wavelength reuse. Besides Rayleigh backscattering (RB) effect, the way to remove the crosstalk caused by residual downstream optical signal-induced upstream signal fluctuations efficiently is a big obstacle in wavelength reused WDM-PONs. There are various methods for removing the crosstalk effect including: a) Using an optical gain saturated SOA-based data eraser to decrease the downstream optical signal before it is re-modulated for transmitting upstream data; b) Making the residual downstream optical signal waveform smooth by injecting feed-forward current in a RSOA intensity modulator [6]. On the other hand, different signal modulations can be exploited for DS and US signals. Among all the signal modulation methods that were discussed, the crosstalk effect imposed on the RSOA intensity modulated US signals can be decreased by the constant downstream FSK and DPSK signal waveforms [6]. On the other hand, broadening the spectrum of the optical signal which is fed by means of wide spectrum modulation formats is another approach to reduce this crosstalk. Enabling efficient coding gain in Rayleigh crosstalk limited systems, forward-error-correction (FEC) codes performance has demonstrated to modify the power budget [25].

A schematic diagram of an RSOA is presented in Fig. 23. The two facets of the SOA are anti-reflection (AR) coated which is employed for the input and output ports and high-reflection (HR) coated or just cleaved for high reflectivity. After amplification and reflecting back to the same port, the seed light is injected to this device.

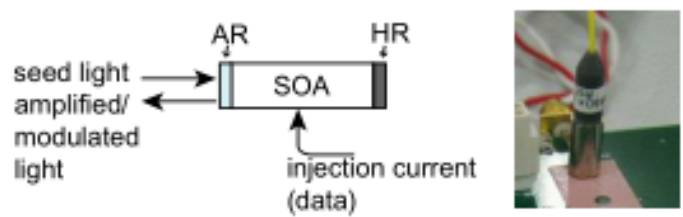

**Figure 23.** Schematic diagram of RSOA [20].

About 10 coarse WDM (CWDM) channels can be covered by recent SOA development. A basic bidirectional single-fiber single-wavelength scenario for a FTTH network is shown in Fig. 24. The transmission wavelength is enabled by a laser stack at the OLT which creates point-to-point connections at each ONU. Also, agile WDM/TDM protocols can create point-to-multipoint connections. An arrayed waveguide grating (AWG) is utilized to obtain wavelength routing at the remote node (RN). DS signal detection and UP data modulation are done by the utilization of a reflective structure at the ONU. In such scheme, the uplink data is imprinted by the RSOA utilizing the downstream optical signal while back-reflecting the signal towards the OLT.

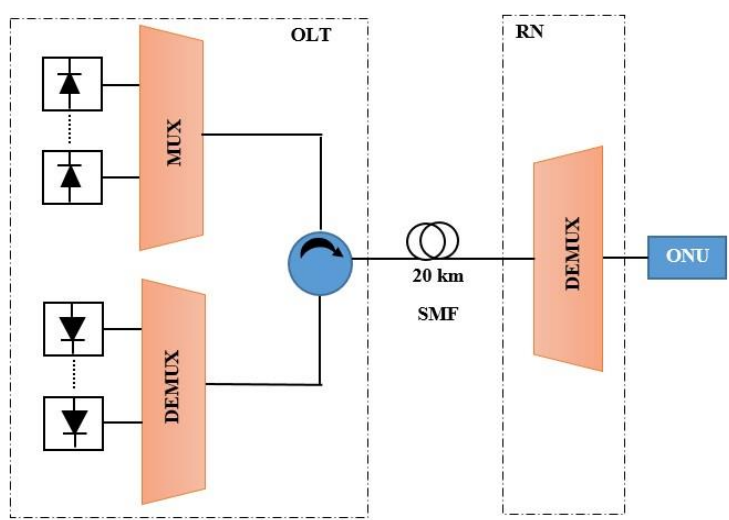

**Figure 24.** Generic bi-directional access network scenario.

Hitherto, three dominant techniques have been mainly experimentally examined to modify the ONU robustness to crosstalk caused by RB. The first one is amplitude shift keying (ASK) data modulation for both UP and DS transmission (ASK-ASK) as shown in Fig. 25. A simple time division with half duplex can be used in this case to prevent overlapping. In order to attain full-duplex communications with a bounded distortion, a constant power offset to the DS modulated signal can be added which is ASK re-modulated by the RSOA. The second one is the combination of frequency shift keying (FSK) for the DS data and ASK for the UP data (FSK-ASK). The third one is utilizing sub-carrier Multiplexing (SCM) in the electrical domain for both UP and DS channels. The second and third systems use full-duplex transmission in the time domain. Also, applying FEC to these systems can enhance the power budget. Instead of utilizing a CW light, exploiting DS signal for re-modulation leads to uplink sensitivity modification. For instance, by the use of a 1.25 Gb/s signal with $2^{23}-1$ PRBS (pseudo-random binary-sequence), a sensitivity of -20 dBm can be attained [25].

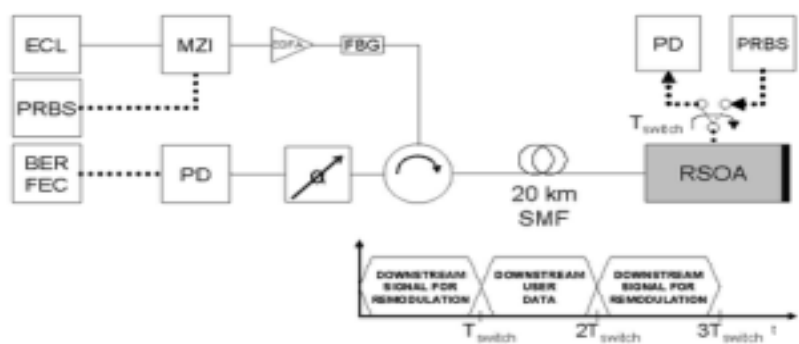

**Figure 25.** Set-up for the ASK-ASK scheme with RSOA.

In FSK-FSK with RSOA (Fig. 26), the DS demodulating section is segregated by the ONU while an optical coupler separates the UP re-modulation section. For the detection of the FSK signal, the transmission frequency corresponding to zero data can be abolished, and the FSK modulation can be converted to intensity modulation by a tunable band-pass filter. For receiving the DS data, a PIN photo-detector is typically utilized. Because of the constant amplitude of the FSK, the RSOA can intense modulate the UP data in the modulation branch with minor distortion.

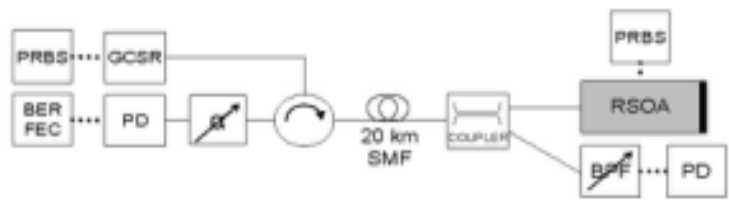

**Figure 26.** Set-up for the FSK-FSK scheme using RSOA.

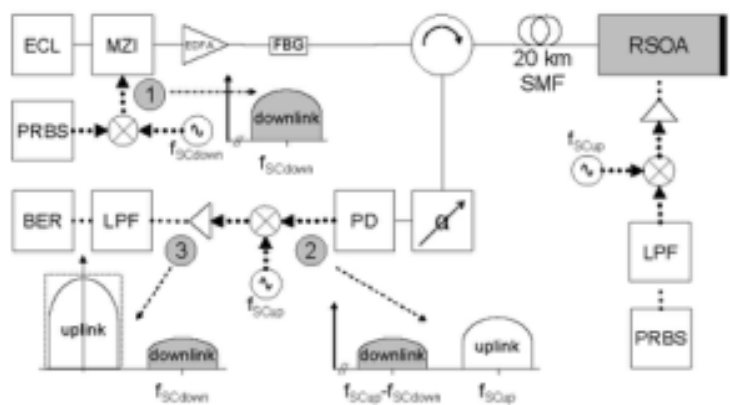

**Figure 27.** Set-up for the SCM scheme using RSOA.

UP and DS signals are multiplexed on different electrical frequencies in SCM with RSOA (Fig. 27). To avoid any crosstalk interference between UP and DS signals, the subcarrier frequency spacing is typically chosen to be small (e.g. 155 MHz). On the other hand, transmitting UP and DS data on various electrical frequency bands can decrease the RB effect. Generally speaking, because of the optical signal detection, modulation and amplification abilities of the SOAs, they play an important role as O/E-devices for implementing ONU [25]. However, being dependent to the temperature with unnecessary uncooled operation is a fundamental obstacle for the cost and reliability of RSOAs. The gain spectra measured at different operating temperatures can be found in Fig. 28.

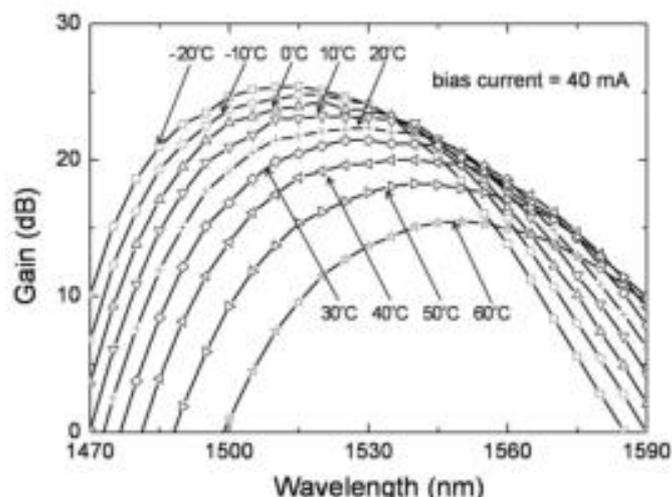

**Figure 28.** Gain spectra of RSOA [20].

There are various techniques for overcoming the demodulation noise, such as the gain-saturated RSOA, Manchester coding and subcarrier multiplexes (SCM). These techniques were utilized for eliminating the demodulation noise instead of the conventional non-return-to-zero (NZR) format for the downstream signals.

RSOA-based WDM PON with a single reflection point is depicted in Fig. 29. Here, the effect of the reflection on the signal when it occurs at the transmission fiber can be categorized as Reflection I and Reflection II. Interfering with the upstream signal while, demonstrating the back-reflected downstream signal has been done by reflection I which produces intensity noises. Reflection II represents the demodulated upstream signal reflected back to the ONU. This can be further divided into two kinds. For first kind, Interfere of the downstream signal with the upstream signal reflected back to the ONU leads to the OBI noise on the downstream signal (Reflection-II (a)). For the second kind, RSOA amplifies this reflected upstream signal again at the ONU and then it merges with the upstream signal (Reflection-II (b)). There have been some assessments for modification of reflection tolerance. It has been shown that broadening the optical spectrum of the signal (e.g., by applying the bias dithering and by phase modulation) can mitigate the reflection-induced OBI noise. Also, utilizing Manchester coding and SCM for the downstream signal has demonstrated to modify the reflection tolerance incredibly. Moreover, employing a low-coherence light (such as ASE light) as the seed light can modify dramatically the reflection tolerance.

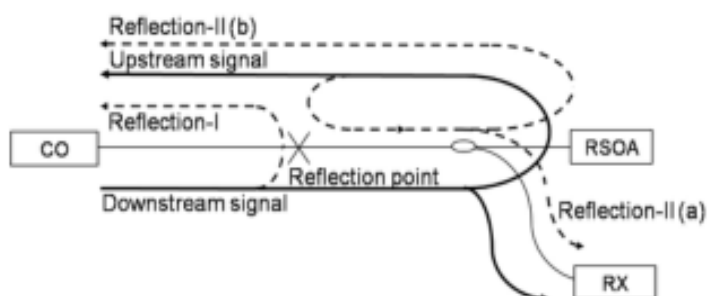

**Figure 29.** Effect of a discrete reflection in RSOA-based WDM PON [20].

## 6. OFDM WDM-PON based on RSOA

Raising the bandwidth in optical access network is essential as very high broadband services grow rapidly. According to FSAN and IEEE discussions, future PON systems deal with a bit rate of 10 Gbit/s. An extensive analysis has been done on Wavelength Division Multiplexing Passive Optical Network (WDM-PON) for future broadband access network. Colorless optical amplifier and modulator at the optical network unit (ONU) play an important role in a WDM-PON for centralizing wavelength management of the channels at the Central Office (CO). A candidate for a low-cost solution in access network has been proposed as utilization of reflective semiconductor optical amplifiers (RSOAs). To have low-cost implementation with conventional low bandwidth components, a high data rate stream can be split into many lower rate sub-streams which are a multi-carrier transmission technique known as AMOOFDM [26]. AMOOFDM has better signal transmission capacity, more network flexibility and more performance robustness than OOFDM [6]. WDM-PON single fiber architecture is shown in Fig. 30.

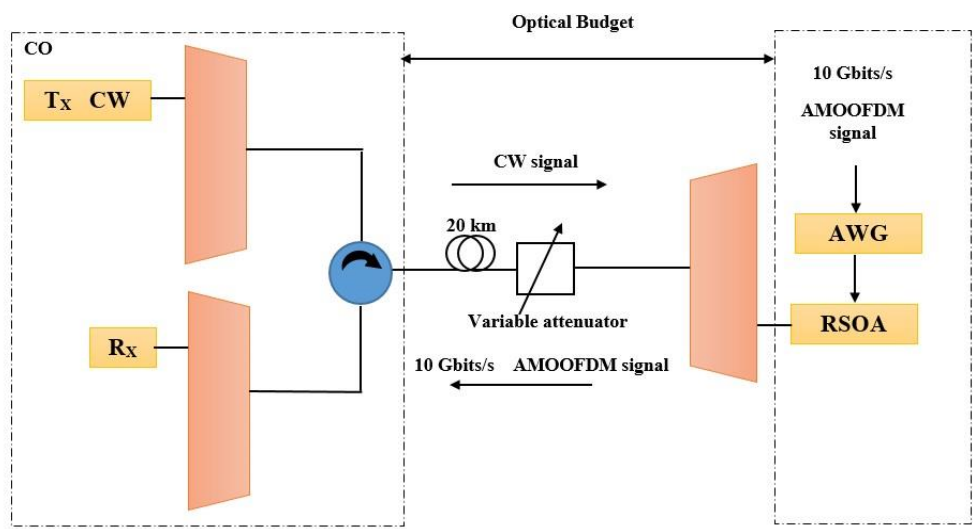

**Figure 30.** Experimental setup.

To increase the requirement for band-width in next generation optical access networks besides gaining longer reach and more capacity than existing PON systems, it is suggested to use WDM-PON. Sharing of optical wavelength and the realization of colorless optical network units (ONU) are implemented to make the WDM-PON system cost effective. They can be provided by reutilization of the downlink signals for uplink transmission which is performed via signal re-modulation. Baseband modulation can produce both the uplink and downlink. Utilizing various modulation formats, cross talk can be minimized for downlink and uplink transmission. Subcarrier multiplexing is a noteworthy candidate for segregating the uplink and the downlink signals into various shares of the available signal spectrum. Single sideband (SSB) modulated subcarrier transmission is preferred for decreasing the effect of fading associated with chromatic dispersion, compared with double sideband (DSB). Utilizing 10 Gb/s OFDM signal is effective for high speed bidirectional signal transmission in a WDM PON system with upper sideband (USB) for downlink transmission and lower sideband (LSB) for uplink transmission. In each optical line terminal (OLT), an intensity modulator (IM) with up converted OFDM signals modulates the optical carrier and DSB signal is produced. Utilizing an optical filter for filtering out one sideband to generate a SSB signal can decrease the power fading effect. After a WDM multiplexer and transmission through the optical distribution network (ODN), consisting of 20km SMF feeder fiber, a WDM multiplexer/demultiplexer and one distribution fiber, the signal is split into two parts. Direct detection of the single sideband OFDM downlink signal utilizes one part while producing the uplink signal by RSOA demodulation of the optical carrier is utilized via the other part. There are several benefits for the system architecture: 1) The power fading effect is decreased by SSB signal for high speed downlink transmission; 2) The modulation of downlink and uplink signals on various frequencies causes the decrease to the effect of RB; 3) Spectral efficiency and transmission bit rate can be increased by utilizing OFDM signals to modulate the bandwidth limited RSOA with high order modulation format; 4) There is no need for more light sources, modulators or optical filters at the ONUs. Managing the Wavelength could be simply obtained and the complex system will be notably simplified. The suggested bidirectional hybrid OFDMWDM PON system architecture is shown in Fig 31.

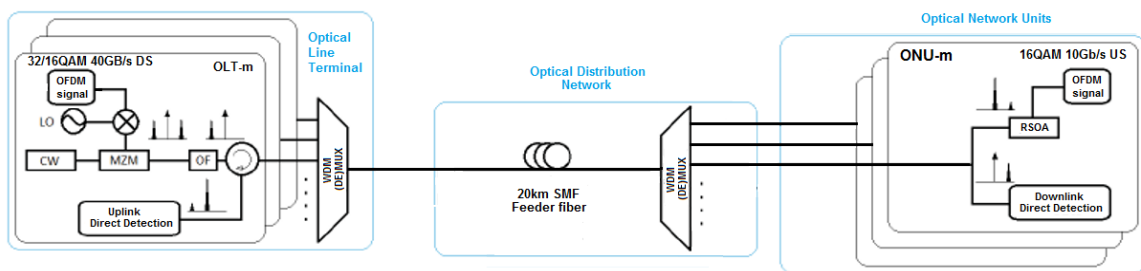

**Figure 31.** Architecture of proposed bidirectional hybrid FDM-WDM-PON.

Because of high spectral efficiency of the OFDM signal, the data rate of the PON can be increased while still utilizing low bandwidth optical components optimized for the present PON. Current commercial RSOA has a typical modulation band width of ~1 GHz. Future WDM-PON desires to use the RSOA-based ONU for > 10 Gbit/s. Some attempts have been done recently for increasing the RSOA operation speed to 10 Gbit/s utilizing modulated optical OFDM (AMOOFDM), offset optical filtering and electronic equalization, special design structure, or adding a delay interferometer. The empirical structure of such WDM-PON is presented in Fig. 32.

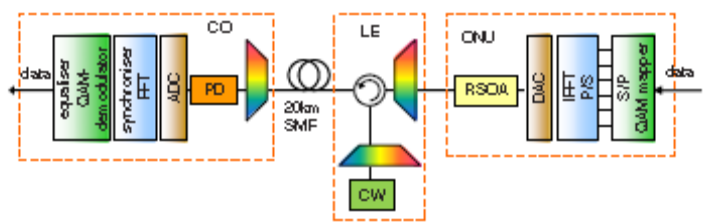

**Figure 32.** Empirical structure of WDM-PON with seeding light source at local exchange (LE) S/P: serial-to-parallel; P/S: parallel-to-serial; IFFT: inverse fast Fourier transform; FFT: fast Fourier transform; DAC: digital-to-analogue converter; ADC: analogue-to-digital converter [28].

As noted in section 6, bold factors bounding the highest accessible downstream and upstream transmission performance are the RB noise effect and crosstalk effect. In order to decrease the crosstalk effect the following steps can be taken: a) Using a gain saturated SOA-based optical data eraser for mitigating the downstream optical signal before re-modulation to transmit upstream data; b) Smoothing the residual downstream optical signal waveform via feed-forward current injection [6] in a RSOA intensity modulator. As another option, various signal modulations can be utilized for obtaining maximum attainable downstream and upstream transmission performances. 10Gb/s downstream and 6Gb/s upstream over 40km of SMF can be achieved via optimum RSOA operating with practical downstream and upstream optical launch powers. Specifically, introducing SSB subcarrier modulation (SSB-SCM) in the downstream systems can lead to 23 Gb/s downstream and 8Gb/s upstream over 40 km SMF. The wavelength-reused bidirectional transmission colorless WDM-PON architecture discussed here is presented in Fig. 33. The modulation format taken on each subcarrier within a symbol ranges from differential binary phase shift keying (DBPSK), differential quadrature phase shift keying (DQPSK), 8-quadrature QAM to 256-QAM. Regarding the downstream OFDM transmitter and downstream OFDM receiver, a high (low) modulation format can be utilized on a subcarrier experiencing a high (low) signal-to-noise ratio (SNR). In case of occurring large number of errors even when using the lowest modulation format, any subcarrier suffering from a very low SNR may be completely dropped. In order to ensure a positive value for each sample, the real-valued electrical OFDM signal from the downstream OFDM transmitter is up-shifted by incorporating an optimized DC bias current. After that, the SOA is straightly driven by the up-shifted electrical OFDM signal so that an injected optical CW wave is modulated via changing the SOA optical gain. At the end, a variable optical attenuator, an optical circulator and a multiplexer couple and multiplex the SOA intensity modulated AMOOFDM downstream signal into a standard SMF. Then after de-multiplexing the received downstream AMOOFDM signal and transmission through the SMF, it is split into two optical beams. Utilization of a square-law photon detector detects the first beam and the data recovery takes place in the downstream OFDM receiver while an inverse procedure is done for the downstream OFDM transmitter. An optical circulator injects the second optical beam into a RSOA intensity modulator. A precious up-shifted and electrically amplified OFDM signal resulted from the upstream OFDM transmitter drives the RSOA intensity modulator. In the RSOA intensity modulator, the upstream electrical OFDM signal changes the RSOA optical gain and this signal re-modulates the injected downstream optical signal. This generated upstream AMOOFDM signal is combined with the same SMF link and then it is compared to the downstream AMOOFDM signal. Then, the combined upstream optical power is fixed by an optical attenuator at a particular level. After upstream transmission through the SMF, a square-law photon detector detects the received upstream AMOOFDM signal and an upstream OFDM receiver recovers it. Its method is the reverse of the corresponding upstream OFDM transmitter.

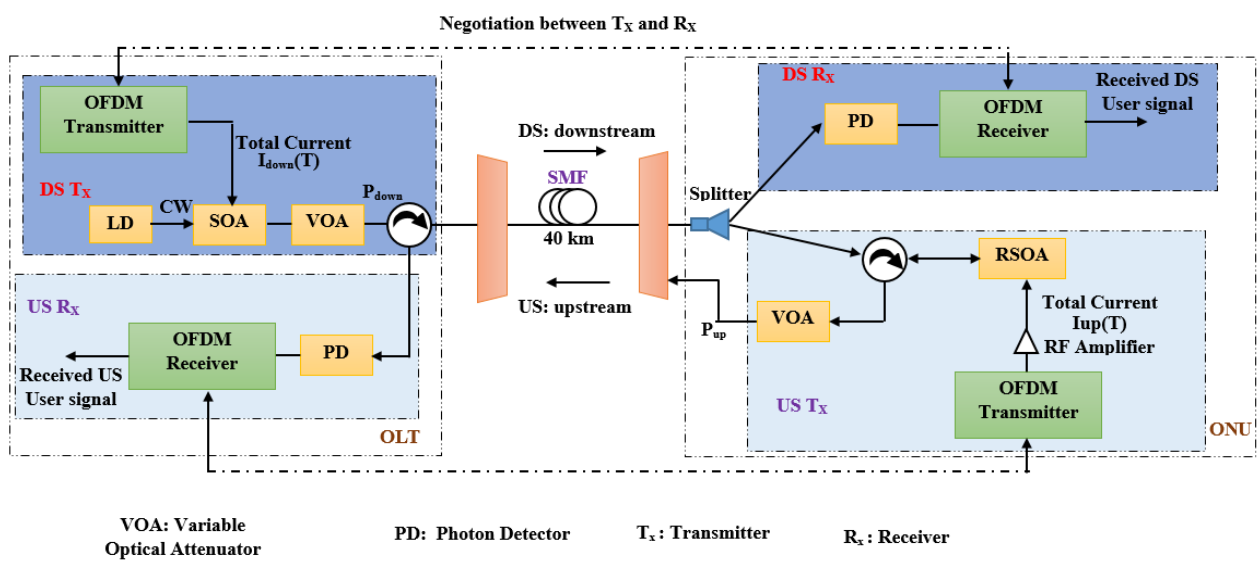

**Figure 33.** Wavelength-reused bidirectional transmission WDM-PON architecture with SOA intensity modulated downstream AMOOFDM signals and RSOA intensity modulated upstream AMOOFDM signals.

According to Fig. 34, two RSOAs at each optical network unit were utilized for the upstream transmission of data at a rate of 2Gb/s in which the system delivers both downstream and upstream on a single wavelength using a CW laser as a pulse source. Exploiting Raman amplifiers at the remote node (RN), the OFDM WDM–PON could reach to 50 km. A different wavelength channel was assigned to each end user by AWGs which can lead to higher bandwidth. Furthermore, from security and expense perspectives, time division multiplexing (TDM) PONs have a considerable role. Recently in a study on OFDM WDM–PON, exploiting a time-domain interleaved OFDM technique, the energy consumption of ONUs in OFDM PONs was demonstrated to be decreased. Here, a CO-OFDM with WDM PON is used. It enables a long-reach OFDM WDM–PON being able to give service to each subscriber that includes both downstream data and upstream data at a rate of 100 Gb/s and 2 Gb/s respectively.

…

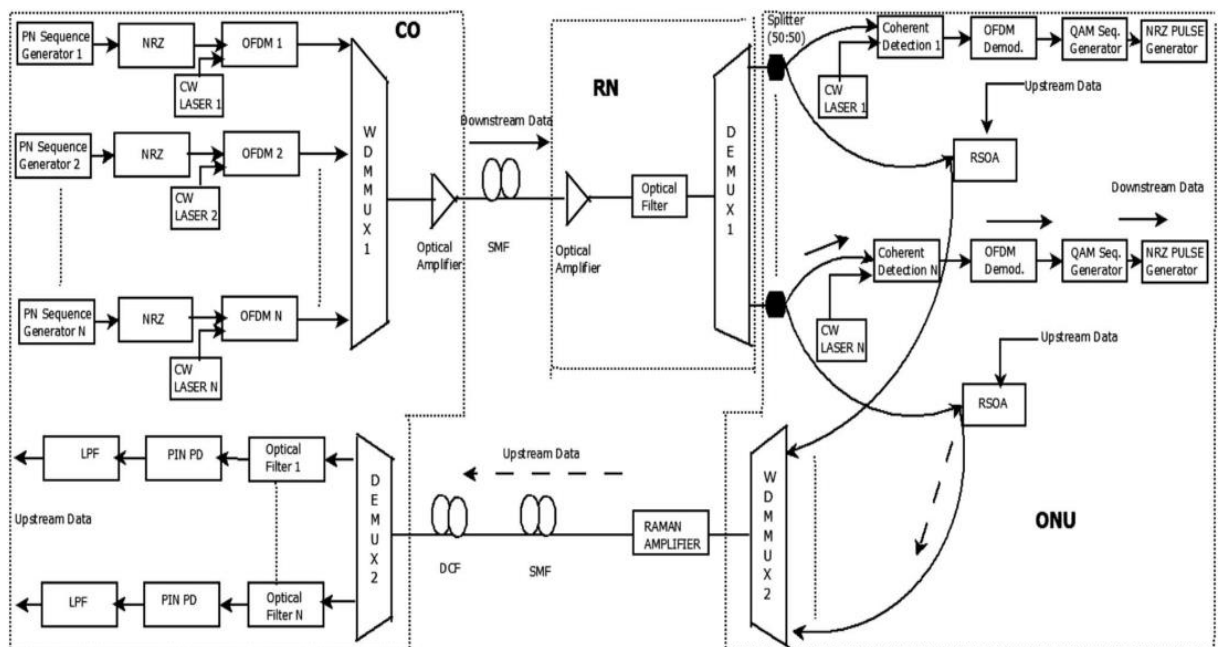

**Figure 34.** Suggested OFDM WDM–PON structure.

By utilizing the multiplexing technique, and incorporating the outputs of two RSOAs at each ONU, it was possible to achieve signal moving upstream at a rate of 2 Gb/s. For secondary signal amplification, the method which incorporates a dispersion compensating fiber with a single mode fiber, a Raman amplifier with an EDFA is employed. Improving the signal power and compensating for the fiber dispersion over a wide wavelength range are motivations for designing this technique. To multiplex a downstream signal of a different wavelength, WDM MUX 1 is employed. The downstream signal feeds into an EDFA and is amplified by it. Then, this amplified downstream signal feeds into single-mode fiber (SMF).

As illustrated in Fig. 35, an EDFA, Bessel optical filter, and a demultiplexer (DEMUX 1) are employed in the RN Downstream. EDFA transmits the signals to enhance the weak signals, making the system capable of working for long-distance use. Then, DEMUX 1 transmits them and demultiplex the downstream signal. After that, the downstream signals are transmitted to an ONU. A splitter splits the optical signal in the ONU. A coherent detector receives half of the optical signal. An RSOA injects the other half of the optical signal to demodulate it with the upstream baseband data. Operating in the gain saturation region, the RSOA can extrude downstream baseband data, and provide the upstream data to be imposed directly upon the downstream signal. Being utilized for multiplexing, WDM MUX 2 incorporates the modulated outputs of these RSOAs again. Then, a Raman amplifier, a SMF and a dispersion-compensating fiber send back the outputs to the central office. After that, DEMUX 2 demultiplex them. Then, the upstream data signal is transmitted to the optical filter. Utilized for receiving the upstream data, a PIN-PD detects output from the optical filter. Afterward, a low-pass filter transmits the output of the PD. The BER for upstream signals can be found at the output of the low-pass filter [29].

## 7. Conclusion

We highlighted the benefits of applying OFDM technology in next-generation (NG) high-capacity multi-channel PONs. We conclude that together with the performance benefits of OFDM modulation such as spectral efficiency, tolerance to both chromatic dispersion and polarization mode dispersion and simple implementation using 1-tap equalization, the utilization of many encoded subcarriers with different modulation formats (adaptive modulation) can result in the optimal use of the available bandwidth (dynamic bandwidth allocation) in NG-PONs. On the other hand, by assigning a specified wavelength to each user can potentially provide increase security in WDM-PONs. An RSOA can also be implemented at the ONU as an affordable element in OFDM-WDM-PON by concurrently amplifying and re-modulating the optical signal, preventing the employment of a separate source. By combining adaptive optical OFDM and RSOA, the signal capacity per user can be maximized for both upstream and downstream PON directions. We believe such configuration could also be very useful to very-high spectral efficient modulation formats such as amplitude shift keying based Fast-OFDM combined with adaptive modulation, which is twice more spectral efficient than conventional OFDM [30-32].